\documentclass[seceq]{ptptex}

\def\ignore#1{{}}

\renewcommand{\thetable}{\Roman{table}}

\newcommand{\beeq}{\begin{equation}}
\newcommand{\eneq}{\end{equation}}
\newcommand{\beqn}{\begin{eqnarray}}
\newcommand{\eeqn}{\end{eqnarray}}

\def\mybig{\displaystyle \strut }

\def\dd{\partial}
\def\la{\raise.16ex\hbox{$\langle$}\lower.16ex\hbox{}  }
\def\ra{\, \raise.16ex\hbox{$\rangle$}\lower.16ex\hbox{} }

\def\next{{~~~,~~~}}
\def\onehalf{ \hbox{${1\over 2}$} }

\def\Tr{{\rm Tr \,}}

\def\eff{{\rm eff}}
\def\H{{\cal H}}

\def\sym{{\rm sym}}

\def\diag{{\rm diag ~}}

\def\GUT{{\rm GUT}}
\def\SUSY{{\rm SUSY}}
\def\B{{\rm B}}

\def\vphi{\varphi}
\def\ep{\epsilon}
\def\psibar{ \psi \kern-.65em\raise.6em\hbox{$-$} }
\def\psibarl{ \psi \kern-.65em\raise.6em\hbox{$-$} \lower.6em\hbox{} }

\title{
Classification and Dynamics of Equivalence Classes\\
in $SU(N)$ Gauge Theory on the Orbifold $S^1/Z_2$%
}

\author{
Naoyuki \textsc{Haba}$^{1}$, Yutaka \textsc{Hosotani}$^{2}$ and
Yoshiharu \textsc{Kawamura}$^{3}$}%

\inst{
$^1${Institute of Theoretical Physics,
 University of Tokushima, Tokushima 770-8502, Japan} \\
$^2${ Department of Physics, Osaka University,
Toyonaka, Osaka 560-0043, Japan}\\
$^3${Department of Physics, Shinshu University,
Matsumoto, Nagano 390-8621, Japan}
}

\abst{
The dynamical determination of  the boundary conditions in
$SU(N)$ gauge theory on the orbifold $S^1/Z_2$ is investigated.   
We classify the equivalence classes of the boundary conditions, and then
the vacuum energy density of the theory in each equivalence class is evaluated
at  one loop order.   Unambiguous  comparison of the vacuum energy densities in
the two theories in different equivalence classes becomes possible in
supersymmetric theories. It is found that in the supersymmetric $SU(5)$
models with the Scherk-Schwarz supersymmetry breaking,  the theory with
the boundary conditions yielding the standard model symmetry is in
the equivalence class with the lowest energy density,
though the low energy theory is not identically the minimal 
supersymmetric standard model.  We also study how
particular boundary conditions are chosen in the  cosmological evolution
of the universe.
}

\begin{document}

\maketitle

\section{Introduction}

In higher-dimensional grand unified theory  (GUT), the gauge fields 
and the Higgs fields in the adjoint representation in the lower 
four-dimensions are unified.\cite{YH1,YH2}  Previously Manton attempted 
to unify  the doublet Higgs fields  with the gauge fields in  the
electroweak  theory by  adopting a monopole-type gauge field configuration
in the extra-dimensional space in a larger gauge group.\cite{Manton1} 
However, such a configuration with nonvanishing field strengths has 
a higher energy density, and  the dynamical stability of the field
configuration  remaines to be justified. It since has been recognized that
in higher-dimensional GUT  defined on a multiply connected manifold,  the
dynamics of Wilson line phases  can induce dynamical gauge symmetry breaking
by developing nonvanishing expectation values through radiative corrections.
This provides real unification of the gauge fields and Higgs fields. 
Here the symmetry is broken by dynamics,  not by hand. This
Hosotani  mechanism has  been extensively investigated since that time,
\cite{WilsonL1}-\cite{WilsonL3} 
but the incorporation of chiral fermions has been a major
obstacle to constructing a realistic model.\cite{Candelas, YH4, chiral}  This
idea has been recently revived under the name of   `gauge-Higgs
unification'.\cite{gaugeHiggs} Further, when the extra dimensions are
deconstructed,  some of the extra-dimensional components of the gauge fields
acquire masses through radiative corrections, and pseudo-Nambu-Goldstone bosons in 
the lower four dimensions   become `little
Higgs'.\cite{littleHiggs}

Significant progress has been made in this direction by considering
GUT on an orbifold.  This provides a new way
to solve such problems  as the chiral fermion problem of how
four-dimensional chiral fermions are generated from a higher-dimensional
space-time  and the Higgs mass splitting problem
\cite{Higgs} (the
$\lq$triplet-doublet mass splitting problem' in $SU(5)$ GUT
\cite{Kawamura1,Hall1}). In formulating gauge theory on an
orbifold,\cite{Hall1}-\cite{HHHK} however, there arise many
possibilities for boundary conditions (BCs) to be imposed on the fields in
the extra-dimensional  space. This leads to  the problem of which type of BCs
should be imposed without relying on  phenomenological information.
We refer to this problem as the $\lq$arbitrariness problem'.\cite{YH3}

The arbitrariness problem is partially solved
at the quantum level by the Hosotani mechanism
as shown in our previous paper.\cite{HHHK}\footnote{
See Refs.\ 26) and 27) for SUSY breaking and
Ref.\ 18) for gauge symmetry breaking on $S^1/Z_2$ by the Hosotani
mechanism.}
The rearrangement of gauge symmetry takes place through the dynamics
of the Wilson line phases.  The physical symmetry
of the theory, in general, differs from the symmetry of the BCs.
Several sets of BCs with distinct
symmetry can be related  by large gauge transformations, belonging to the
same equivalence class.  This implies the reduction of   the number of
independent theories.

The remaining problem  is to determine how a realistic theory is selected  among
various equivalence classes.  We have not obtained a definite solution to this
problem, partially because of a lack of  understanding of a more
fundamental theory that contains dynamics which select various BCs.
Nevertheless,  one can expect  that the classification and characterization of
equivalence classes  will provide information as to how one of them is
selected dynamically.  In particular, the evaluation of the vacuum energy
density in each equivalence class would provide  critical information  
needed to solve the arbitrariness problem.\cite{YH3}   This   motivates the
present work.

In the present paper, we classify the equivalence classes of the BCs in
$SU(N)$ gauge theory on $S^1/Z_2$, and evaluate the vacuum energy density
in each equivalence class.
Some discussion is given with regard  to the question of how particular BCs are
selected in the cosmological evolution of the universe.

In \S 2    general arguments are given for BCs in gauge theories on
the orbifold $S^1/Z_2$, and
we classify those BCs  with the aid of equivalence relations originating
from the gauge invariance.
In \S 3  generic formulas for  the one-loop effective
potential at vanishing Wilson line phases are derived.
These can be applied to every equivalence class.
The arguments are generalized to  supersymmetric theories in \S 4.
There it is recognized that unambiguous comparison of the
vacuum energy densities in theories on different equivalence classes
is possible.
In \S 5, we discuss how particular boundary conditions
are chosen in the  cosmological evolution of the universe.
Section 6 is devoted to conclusions and discussion.

\section{Orbifold conditions and classification of equivalence classes}

In this article, we focus on $SU(N)$ gauge theory defined
on a five-dimensional space-time $M^4 \times (S^1/Z_2)$
where $M^4$ is the four-dimensional Minkowski spacetime.
The fifth dimension, $S^1/Z_2$, is obtained by identifying
two points on $S^1$ by parity.  Let $x$ and $y$ be coordinates of
$M^4$ and $S^1$, respectively.   $S^1$ has a radius $R$, and therefore
a point $y+2\pi R$ is identified with a point $y$.
The orbifold $S^1/Z_2$ is obtained by further identifying
$-y$ and $y$.  The resultant fifth dimension is the
interval $0 \le y \le \pi R$, which contains the information on $S^1$.

\subsection{ Boundary conditions}

For $Z_2$ transformations around $y=0$ and $y = \pi R$ and
a loop translation along $S^1$, each defined by
\beqn
Z_0: ~ y \to -y, ~~~ Z_1: ~ \pi R + y \to \pi R - y, ~~~ S: ~ y \to y + 2\pi
R ~~,
\label{Z2S}
\eeqn
the following relations hold:
\beqn
Z_0^2 = Z_1^2 = I , ~~~ S = Z_1 Z_0 , ~~~ S Z_0 S = Z_0 ~~.
\label{Z2S-rel}
\eeqn
Here $I$ is the identity operation.
Although we have the identification $y \sim y + 2 \pi R \sim -y$ on
$S^1/Z_2$,
fields do not necessarily take identical values
at $(x, y)$, $(x, y + 2\pi R)$ and $(x, -y)$
as long as  the Lagrangian density   is single-valued.
The general BCs for a field $\vphi(x,y)$ are given by
\beqn
&~& \vphi(x, -y) = T_{\vphi}[P_0] \vphi(x, y) , ~~~
\vphi(x, \pi R -y) = T_{\vphi}[P_1] \vphi(x, \pi R + y) ,
\nonumber \\
&~& \vphi(x, y + 2 \pi R) = T_{\vphi}[U] \vphi(x, y) ,
\label{BCs-vphi}
\eeqn
where $T_{\vphi}[P_0]$, $T_{\vphi}[P_1]$ and $T_{\vphi}[U]$
represent appropriate representation matrices, including an arbitrary  sign
factor.  The counterparts of (\ref{Z2S-rel}) are given by
\beqn
T_{\vphi}[P_0]^2 = T_{\vphi}[P_1]^2 = I , ~~
T_{\vphi}[U] = T_{\vphi}[P_0] T_{\vphi}[P_1] ,
~~ T_{\vphi}[U]T_{\vphi}[P_0]T_{\vphi}[U] = T_{\vphi}[P_0] .
\label{vphi-rel}
\eeqn
In (\ref{vphi-rel}) $I$ represents an unit matrix.
For instance, if $\vphi$ belongs to the
fundamental representation of the $SU(N)$ gauge group,
then $T_\vphi [P_0] \vphi$ is $\pm P_0 \vphi$ where $P_0$ is
a hermitian $U(N)$ matrix, i.e., $P_0^\dagger = P_0 = P_0^{-1}$.
The same property applies to $P_1$.

The BCs  imposed on a gauge field $A_M$ are
\beqn
&&\hskip -1cm
A_\mu (x, - y ) = P_0 A_\mu(x,y) P_0^\dagger , ~ A_y (x, - y ) = - P_0 A_y
(x,y) P_0^\dagger ,
\label{OrbiBC1}\\
&&\hskip -1cm
A_\mu (x, \pi R - y ) = P_1 A_\mu(x,\pi R + y) P_1^\dagger , ~
A_y (x, \pi R - y ) = - P_1 A_y (x,\pi R + y) P_1^\dagger ,
\label{OrbiBC2}\\
&&\hskip -1cm
A_M (x, y + 2\pi R) = U A_M(x,y) U^\dagger .
\label{S1BC1}
\eeqn
The BCs of scalar fields $\phi^A$ are given by
\beqn
&&\hskip -1cm
\phi^A(x,-y) = T_{\phi^A}[P_0] \phi^A(x,y) ~~,\cr
\noalign{\kern 10pt}
&&\hskip -1cm
\phi^A(x, \pi R -y)
= \sum_B {\big( e^{i\pi \beta M} \big)^A}_B T_{\phi^B}[P_1] 
  \phi^B(x,\pi R + y) ~~, \cr
\noalign{\kern 3pt}
&&\hskip -1cm
\phi^A(x, y + 2\pi R)
= \sum_B {\big( e^{i\pi \beta M} \big)^A}_B T_{\phi^B}[U] \phi^B (x,y) ~~ ,
\label{OrbiBC3}
\eeqn
where $A$ and $B$ are indices and $M$ is a matrix in the
flavor space. If nontrivial $Z_2$ parity is
assigned in the flavor space and $M$ anti-commutes with the $Z_2$ parity,
then $\beta$ can take an arbitrary value.
For Dirac fields $\psi^A$ defined in the bulk, the gauge invariance of the
kinetic
energy term requires
\beqn
&&\hskip -1cm
\psi^A(x,-y) = T_{\psi^A}[P_0] \gamma^5 \psi^A(x,y) ~~, \cr
\noalign{\kern 10pt}
&&\hskip -1cm
\psi^A(x, \pi R -y) =  \sum_B {\big( e^{i\pi \beta M} \big)^A}_B
T_{\psi^B}[P_1]
  \gamma^5 \psi^B(x,\pi R + y)  ~~, \cr
\noalign{\kern 3pt}
&&\hskip -1cm
\psi (x, y + 2\pi R)
= \sum_B {\big( e^{i\pi \beta M} \big)^A}_B T_{\psi^B}[U] \psi^B(x,y) ~~.
\label{OrbiBC4}
\eeqn

To summarize, the BCs in gauge theories on $S^1/Z_2$ are
specified with
$(P_0, P_1, U, \beta)$ and additional sign factors.
The matrices $P_0$ and $P_1$ need not be diagonal in general.  One can
always diagonalize one of them, say $P_0$, through a global gauge
transformation, but  $P_1$ might not be diagonal.  For the reasons described
in the subsequent sections, we  consider BCs with diagonal $P_0$ and
$P_1$.

The diagonal $P_0$ and $P_1$ are specified by three non-negative integers
$(p, q, r)$ such that
\beqn
&&\hskip -1cm
\diag P_0 = (\overbrace{+1, \cdots, +1, +1, \cdots, +1 ,
        -1, \cdots, -1, -1, \cdots, -1}^N) ~, \nonumber \\
&&\hskip -1cm
\diag P_1 = (\underbrace{+1, \cdots, +1}_{p}, \underbrace{-1,
\cdots, -1}_{q} ,
\underbrace{+1, \cdots, +1}_r, \underbrace{-1, \cdots, -1}_{s = N-p-q-r}) ~,
\label{pqr}
\eeqn
where $N \geq p, q, r, s \geq 0$.
We denote each BCs specified by $(p, q, r)$ (or a theory with such BCs)
as $[p; q, r; s]$.
The matrix $P_0$ is interchanged with $P_1$ by the interchange of
$q$ and $r$ such that
\beeq
[p; q, r; s] \leftrightarrow [p; r, q; s] .
\label{interchange}
\eneq

\subsection{Residual gauge invariance and equivalence classes}

Given the BCs $(P_0, P_1, U, \beta)$, there
still remains residual gauge invariance.  Recall that under a
gauge transformation $\Omega(x,y)$, we have
\beqn
&~& A_M \to  {A'}_M = \Omega A_M  \Omega^{\dagger}
   -  {i \over g}\Omega  \partial_M \Omega^{\dagger} ~~,  \cr
&~& \phi^A \to
  {\phi'}^A = T_{\phi^A} [\Omega] \phi^A ~~,  ~~
\psi^A \to {\psi'}^A = T_{\psi^A}[\Omega] \psi^A  ~~,
\label{gauge1}
\eeqn
where $g$ is a gauge coupling constant.
The new fields $A'_M$ satisfy, instead of
(\ref{OrbiBC2}) -- (\ref{S1BC1}),
\beqn
\begin{pmatrix}
A_\mu' (x, - y) \cr  A_y' (x, - y)
\end{pmatrix}
&=& P_0' 
\begin{pmatrix}
 A_\mu' (x,y) \cr - A_y' (x,y) 
\end{pmatrix}
 P_0'^\dagger
   - {i\over g} \,  P_0' \,
\begin{pmatrix}
 \dd_\mu \cr - \dd_y 
\end{pmatrix}
 P_0'^\dagger ~~, \cr
\noalign{\kern 10pt}
\begin{pmatrix}
A_\mu' (x,\pi R- y) \cr A_y' (x,\pi R- y) 
\end{pmatrix}
&=& P_1' 
\begin{pmatrix}
A_\mu' (x,\pi R + y) \cr -A_y' (x,\pi R + y) 
\end{pmatrix}
  P_1'^\dagger
   - {i\over g} \,  P_1' \,
\begin{pmatrix}
 \dd_\mu \cr -\dd_y
\end{pmatrix}
  P_1'^\dagger ~~, \cr
\noalign{\kern 10pt}
A_M' (x, y + 2\pi R) &=& U' A_M'(x,y) U'^\dagger
   - {i\over g} U' \dd_M U'^\dagger ~~, 
\label{BC2}
\eeqn
where
\beqn
&&\hskip -1cm
P_0' = \Omega(x,-y) \, P_0 \, \Omega^\dagger (x,y) ~~, ~~
P_1' = \Omega(x,\pi R -y) \, P_1 \, \Omega^\dagger (x,\pi R + y)  ~~, \cr
\noalign{\kern 5pt}
&&\hskip -1cm
U' = \Omega(x,y+2\pi R) \,  U \, \Omega^\dagger (x,y) ~~.
\label{BC3}
\eeqn
Other fields ${\phi'}^A$ and ${\psi'}^A$ satisfy relations similar to
(\ref{OrbiBC3}) and (\ref{OrbiBC4}) where
$(P_0, P_1, U)$ are replaced by  $(P_0', P_1', U')$.

The residual gauge invariance of the BCs is given by gauge
transformations  which preserve the given BCs, namely those
transformations which satisfy $U'=U$, $P_0'=P_0$ and $P_1'=P_1$:
\beqn
&&\hskip -1cm
\Omega(x,-y) \, P_0  = P_0 \,  \Omega (x,y) ~~, ~~
\Omega(x,\pi R -y) \, P_1 = P_1  \Omega (x,\pi R + y)  ~~, \cr
&&\hskip -1cm
\Omega(x,y+2\pi R) \,  U  =  U \, \Omega (x,y) .
\label{residual1}
\eeqn
We call the residual gauge invariance of the BCs
the symmetry of the BCs.

The low energy symmetry of the BCs which is defined independently of the
$y$-coordinate,  is given by
\beqn
&&\hskip -1cm
\Omega(x) \, P_0  = P_0 \,  \Omega (x) ~~, ~~
\Omega(x) \, P_1 = P_1  \Omega (x)  ~~,~~
\Omega(x) \,  U  =  U \, \Omega (x) ~~,
\label{residual2}
\eeqn
that is, the symmetry is generated by generators that commute with
$P_0$, $P_1$ and $U$.

Theories with different BCs can be equivalent with regard to physical
content. The key observation is that in gauge theory, physics should not
depend on the gauge chosen so that one is always free to choose the gauge.
If $(P_0', P_1', U')$ satisfies the conditions
\beqn
&&\hskip -1cm
\dd_M P_0' = 0 ~~,~~ \dd_M P_1' = 0 ~~, ~~\dd_M U' = 0 ~~,~~
P_0'^\dagger = P_0' ~~,~~ P_1'^\dagger = P_1' ~~,
\label{BC4}
\eeqn
then the two sets of the BCs are equivalent:
\beeq
(P_0', P_1', U') \sim (P_0, P_1,U) ~~.
\label{equiv1}
\eneq
It is easy to show that $(P_0', P_1', U')$ satisfy the relations
(\ref{vphi-rel}).
The equivalence relation (\ref{equiv1})
defines equivalence classes of the BCs.
Here we stress that the BCs indeed change under general gauge
transformations.
To illustrate this, let us consider an $SU(2)$ gauge theory with
$(P_0, P_1, U)=(\tau_3, \tau_3, I)$.
If we carry out the gauge transformation
$\Omega = \exp \{ i(\alpha y/2 \pi R) \tau_2 \}$, we find the equivalence
\beeq
(\tau_3, \tau_3, I) \sim
(\tau_3,  e^{i\alpha \tau_2} \tau_3, e^{i\alpha \tau_2}) ~~.
\label{equiv2}
\eneq
In particular, for  $\alpha = \pi$ we have the equivalence
\beeq
(\tau_3, \tau_3, I) \sim
(\tau_3,  -\tau_3, -I) ~~.
\label{equiv2-1}
\eneq
Using this equivalence, we can derive the following equivalence relations in
$SU(N)$ gauge theory:
\beqn
[p; q, r; s] &\sim& [p-1; q+1 , r+1; s-1] , ~~ \mbox{for} ~~ p, s  \geq 1
~~,
\nonumber \\
&\sim& [p+1; q-1 , r-1; s+1] , ~~ \mbox{for} ~~ q, r  \geq 1 .
\label{equ-rel}
\eeqn
The symmetry of the BCs in one theory differs from that in the other, but
the
two theories are  connected by the BCs-changing gauge transformation and
are equivalent. This equivalence is guaranteed by the Hosotani mechanism as
explained in the next subsection.

\vskip .5cm
\subsection{The Hosotani mechanism and physical symmetry}

The two theories with distinct symmetry of BCs are equivalent to each other
in physics content.
This statement is verified by the dynamics of the Wilson line phases
as a part of the Hosotani mechanism.
The Hosotani mechanism in gauge theories defined on multiply
connected manifolds consists of several parts.\cite{YH1,YH2}

\noindent  (i)  Wilson line phases are phases of $WU$ defined by
\beeq
W U \equiv P \exp \Big\{ ig  \int_C dy A_y \Big\} U ,
\label{Wilson}
\eneq
where
$C$ is a non-contractible loop.
The  eigenvalues of $WU$ are gauge invariant and become
physical degrees of freedom.
Wilson line phases cannot be gauged away and parametrize
degenerate vacua at the classical level.

\noindent (ii) The degeneracy is lifted by quantum effects in general.
The physical vacuum is given by
the configuration of Wilson line phases that minimizes
the effective potential $V_\eff$.

\noindent (iii) If the configuration of the Wilson line phases is
non-trivial,
the gauge symmetry is spontaneously broken or restored by radiative
corrections.
Nonvanishing expectation values of the Wilson line phases
give masses to those gauge fields in lower dimensions whose gauge
symmetry is broken.
Some of matter fields also acquire masses.

\noindent (iv)  A nontrivial $V_\eff$ also implies that all
extra-dimensional components of gauge fields become massive.

\noindent (v) Two sets of BCs for fields can be related
to each other by a BCs-changing gauge transformation.
They are physically equivalent, even if the two sets have distinct
symmetry of the BCs.  This defines equivalence classes of
the BCs.   $V_\eff$ depends on the BCs so that
the expectation values of the Wilson line phases depend on the
BCs.  The physical symmetry of the theory is determined
by the combination of the BCs and the expectation values
of the Wilson line phases.  Theories in the same equivalence class of the
BCs have the same physical symmetry and physical content.

\noindent (vi) The physical symmetry of the theory is mostly dictated by the
matter content of the theory.

\noindent (vii)  The mechanism provides unification of gauge fields
and Higgs scalar fields in the adjoint representation, namely
the gauge-Higgs unification.

Let us spell out the  part (v) of the mechanism in gauge theory defined on
$M^4 \times (S^1/Z_2)$.
Dynamical  Wilson line phases are given by
$\{ g \pi R A_y^a ~,~ \onehalf \lambda^a \in \H_W \}$, 
where $\H_W$ is a set of generators which anti-commute
with $P_0$ and $P_1$:
\beeq
\H_W = \Bigg\{ ~{\lambda^a\over 2} ~;~
\{ \lambda^a, P_0 \} = \{ \lambda^a, P_1 \} =0 ~
\Bigg\} ~~.
\label{decomposition2}
\eneq
Suppose that with $(P_0, P_1, U, \beta)$,   $V_\eff$ is minimized
at $\la  A_y \ra$ such that $W \equiv \exp(i 2\pi g R  \la A_y \ra) \neq I$.
Perform a BCs-changing gauge transformation given by
$\omega = \exp\{i \pi g (y + \alpha) \la A_y \ra\}$.
This transforms $\la A_y \ra$ into $\la {A'}_y \ra = 0$.
Under this transformation,  the BCs change to
\beeq
(P_0^\sym, P_1^\sym, U^\sym, \beta)
\equiv (P'_0, P'_1, U', \beta) = (e^{2ig\alpha\langle A_y \rangle} P_0,
e^{2ig(\alpha + \pi R)\langle A_y \rangle} P_1, WU, \beta) .
\label{equiv3}
\eneq
Because the expectation values of $A_y'$ vanish in the new
gauge, the physical symmetry is spanned by the generators
that commute with $(P_0^\sym, P_1^\sym, U^\sym)$:
\beeq
\H^\sym =  \Bigg\{ ~{\lambda^a\over 2} ~;~
[ \lambda^a, P_0^\sym ] = [ \lambda^a, P_1^\sym ] =0 ~
\Bigg\} ~~.
\label{decomposition3}
\eneq
The group  generated by $\H^\sym$, $H^\sym$, is the unbroken
physical symmetry of the theory.

\subsection{Classification of equivalence classes}

The classification of equivalence classes of the BCs is reduced to the
classification of $(P_0, P_1)$.  As briefly mentioned in   \S 2.1, 
$P_0$ can be made diagonal through a suitable global gauge
transformation.
Then $P_1$ is not diagonal in general.  As explained in \S 2.3
two BCs, $(P_0, P_1)$ and $(P_0', P_1')$, can be in the same equivalence
class.  In each equivalence class the vacuum with the lowest energy density
is  chosen by the dynamics of Wilson line phases.  Each equivalence
class is characterized by $(P_0^\sym, P_1^\sym)$ in (\ref{equiv3}).

Let $(P_0, P_1)$ be said to be diagonal if both $P_0$ and $P_1$ are
diagonal.  There are three questions of physical relevance;

\noindent
(Q1)  Does each equivalence class have a diagonal representative
$(P_0, P_1)$?

\noindent
(Q2)  Can we choose $(P_0^\sym, P_1^\sym)$ to be   diagonal?

\noindent
(Q3)  Which $(P_0^\sym, P_1^\sym)$, among all equivalence classes,
 has the lowest energy density?

The answer to (Q1) is affirmative.  The proof is given in Appendix A.
It is shown there that nontrivial BCs are reduced to BCs in
$SU(2)$ subspaces.  As for (Q2), we do not have a satisfactory answer at
the moment.   In a previous paper \cite{HHHK},  it is found that in
many of the BCs of physical interest, $(P_0^\sym, P_1^\sym)$ is diagonal.
Even if one  starts with a non-diagonal $(P_0, P_1)$, the Hosotani
mechanism  yields a diagonal $(P_0^\sym, P_1^\sym)$.
If the answer to (Q2)  is affirmative, the investigation of (Q3) becomes
feasible.  What is  to be done is (1) to list all diagonal pairs $(P_0,
P_1)$ and (2) to evaluate the energy density or the effective potential
at the vanishing Wilson line phases in each diagonal $(P_0, P_1)$.

One can  show that the number of equivalence classes  of
BCs is $(N+1)^2$ in the $SU(N)$ model. The proof goes as follows.
We count the number $n_1$  of diagonal pairs $(P_0, P_1)$ and the number
$n_2$  of equivalence relations among those diagonal $(P_0, P_1)$.
As proved in Appendix A, every equivalence class has a diagonal
representative $(P_0, P_1)$.  Hence, the number of equivalence classes
is given by $n_1 - n_2$.

The number $n_1$ is found by counting all the different
$[p;q,r;s]$ ($p+q+r+s=N$) defined in (\ref{pqr}).
We write $[p;q,r;s] = [N-k; q,r; k-j]$, where $j=q+r$.
The value $k$ runs from 0 to $N$, while $j$ runs from 0 to $k$.
Given $(k,j)$, there are $(j+1)$ combinations for $(q,r)$.
Hence,
\beeq
n_1 = \sum_{k=0}^N \sum_{j=0}^k (j+1)
= {1\over 6} (N+1)(N+2)(N+3) ~~.
\label{count1}
\eneq
The equivalence relation (\ref{equ-rel}) is written as
$[N-k; q,r; k-j] = [N-k-1; q+1,r+1; k-j-1]$.  In this case, $k$ runs
from 1 to $N-1$, while $j$ runs from 0 to $k-1$.  Therefore
\beeq
n_2 = \sum_{k=1}^{N-1} \sum_{j=0}^{k-1} (j+1)
= {1\over 6} (N-1)N(N+1) ~~.
\label{count2}
\eneq
Thus the number of equivalence classes is $n_1 - n_2 = (N+1)^2$.

\section{Effective potential}

There exist $(N+1)^2$  equivalence classes in $SU(N)$ gauge theory
on $M^4 \times (S^1/Z_2)$.    We examine the question of which equivalence
class has the lowest energy density.  It may well be that such an
equivalence class is preferentially chosen by the dynamics governing BCs,
the arbitrariness problem thus being solved.
Other scenarios are  possible, however, in the cosmological evolution of
the universe.   We  come back to this point  in \S 5.

We  evaluate the one-loop effective potential $V_\eff$ for each theory
with diagonal $(P_0, P_1)$.
The effective potential $V_\eff$ depends  not only on Wilson line phases but
also on BCs, i.e., $V_\eff = V_\eff [A^0_M; P_0, P_1, \beta]$
where $A^0_M$ is a background configuration of the gauge field $A_M$,  in
$SU(N)$ gauge theory.  In a more fundamental theory,
the BCs $(P_0, P_1)$ would not be  parameters at our disposal, but would be
determined by  dynamics.
The resultant effective theory belongs to
a specific equivalence class of BCs.

Our goal is to find the global minimum of $V_\eff$.
This is not an easy task,  as it is difficult to write down
a generic formula for $V_\eff$ including $A^0_M$ explicitly.
We consider the case of vanishing VEV's of $A_M$
with diagonal $(P_0, P_1)$.  As remarked in the previous section,
the global minimum of $V_\eff$  in many cases of physical interest
has been found at diagonal $(P_0^\sym, P_1^\sym)$.

The effective potential for $A^0_M$ is derived by writing
$A_M = A^0_M + A^q_M$, taking a suitable gauge fixing and integrating over
the quantum part $A^q_M$.
If the gauge fixing term is also invariant under the gauge
transformation, i.e.,
\begin{eqnarray}
D^{M}(A^{0}) A_M = 0 \to D^{M}({A'}^{0}) A'_M = \Omega D^{M}(A^{0}) A_M
\Omega^{\dagger} = 0 ~~,
\label{gauge-inv-gf}
\end{eqnarray}
it is shown that $V_\eff$ on $M^4 \times (S^1/Z_2)$ satisfies the relation
\begin{eqnarray}
V_\eff [A^0_M; P_0, P_1, U, \beta]
= V_\eff [A'^0_M; P_0', P_1', U', \beta] ~~.
\label{Veff3}
\end{eqnarray}
This property implies that the minimum of $V_\eff$ corresponds to the same
symmetry as that of $(P_0^\sym, P_1^\sym, U^\sym)$.

The one-loop effective potential for $A_M^0$ on $M^4 \times (S^1/Z_2)$ is
given by
\begin{eqnarray}
V_\eff [A^0_M; P_0, P_1, U, \beta] &=& \sum \mp {i \over 2} \, \Tr \ln
D_M(A^0) D^M(A^0)  ~~,
\label{Veff1}\\
 &=& \sum \mp {i \over 2} \int {d^4 p \over (2 \pi)^4}
  ~{1 \over  \pi R} ~ \sum_{n} \ln (- p^2 +  M_n^2 - i \varepsilon) ~~,
\label{Veff2}
\end{eqnarray}
where we have supposed that $F_{MN}^0 = 0$ and every scalar field is also
massless.
The sums extend over all degrees of freedom of fields in the bulk in
(\ref{Veff1})
and all degrees of freedom of 4-dimensional fields whose masses are $M_n$ in
(\ref{Veff2}).
The sign is negative (positive) for bosons (FP ghosts and fermions).
$D_M(A^0)$ denotes an appropriate covariant derivative with respect
to $A^0_M$. The quantity $V_\eff$ depends on $A^0_M$
and the BCs,  $(p,q,r;\beta)$.
Hereafter we take $A^0_M = 0$ on the basis of the assumption  that $V_\eff$
has a minimum there when both $P_0$ and $P_1$ are taken in an
appropriate diagonal form.

On the orbifold $S^1/Z_2$  all fields are classified  as either
$Z_2$ singlets or  $Z_2$ doublets.
The mode expansion of  $Z_2$ singlet fields $\phi^{(P_0 P_1)}(x,y)$ is
given by
\beqn
&&\hskip -1cm
\phi^{(++)} (x,y) = {1\over \sqrt{\pi R}} ~ \phi_0 (x)
  + \sqrt{{2\over \pi R}}\sum_{n=1}^\infty \phi_n (x) ~
    \cos {ny\over R}  ~~,  \cr
\noalign{\kern 10pt}
&&\hskip -1cm
\phi^{(--)} (x,y) = \sqrt{{2\over \pi R}}
   \sum_{n=1}^\infty \phi_n (x) ~ \sin {ny\over R} ~~, \cr
\noalign{\kern 10pt}
&&\hskip -1cm
\phi^{(+-)} (x,y) = \sqrt{{2 \over \pi R}}
   \sum_{n=0}^\infty \phi_n (x) ~ \cos {(n+ \onehalf) y\over R} ~~,  \cr
\noalign{\kern 10pt}
&&\hskip -1cm
\phi^{(-+)} (x,y) = \sqrt{{2 \over \pi R}}
   \sum_{n=0}^\infty \phi_n (x) ~ \sin {(n+ \onehalf) y\over R}  ~~,
\label{expansion}
\eeqn
where $\pm$ indicates the eigenvalue $\pm 1$ of $Z_2$ parity.
The mass terms in four-dimensional space-time are derived from the
kinetic $y$-derivative terms after compactification.  They are
\beeq
 \left({n \over R}\right)^2 \quad (n \ge 0)
\next
 \left({n \over R}\right)^2 \quad (n \ge 1)
\next
\left({n+\frac{1}{2} \over R}\right)^2  \quad (n \ge 0)
\label{massterms}
\eneq
for $\phi^{(++)}$, $\phi^{(--)}$, and
$\phi^{(+-)} ~/~ \phi^{(-+)}$, respectively.
Let us define $N^{(P_0 P_1)} = N^{(P_0 P_1)}_B - N^{(P_0 P_1)}_F$, 
where $N^{(P_0 P_1)}_B$  ($N^{(P_0 P_1)}_F$) is the number of bosonic
(fermionic) fields whose $Z_2$ parities are $P_0$ and $P_1$.
The value $N^{(P_0 P_1)}$ depends on $(p,q,r)$ of the BCs.
For $Z_2$ singlet fields,  the formula for $V_\eff$ becomes
\beeq
V_\eff|_{Z_2 ~{\rm singlet}}
= N_0 \ep_0 + N_\Delta \Delta \ep + N_v v(\onehalf)
\label{V1}
\eneq
where
\beqn
&&\hskip -1cm
N_0 = N^{(++)} + N^{(--)} + N^{(+-)} + N^{(-+)} ~~,
\label{N0} \\
&&\hskip -1cm
N_\Delta = N^{(++)} - N^{(--)} ~~,
\label{NDelta} \\
&&\hskip -1cm
N_v =  N^{(+-)} + N^{(-+)} ~~,
\label{Nv} \\
&&\hskip -1cm
\ep_0 \equiv -\frac{1}{4} \int {d^4 p_E \over (2 \pi)^4}
  ~{1 \over  \pi R} ~ \sum_{n=-\infty}^{\infty} \ln
\left[p_E^2 +  \left(\frac{n}{R}\right)^2 \right] ~~,
\label{ep} \\
&&\hskip -1cm
\Delta \ep \equiv -\frac{1}{4} \int {d^4 p_E \over (2 \pi)^4}
  ~{1 \over  \pi R}  \ln
\left[p_E^2 \right] ~~,
\label{Dep} \\
&&\hskip -1cm
 v(\beta) \equiv - \frac{1}{4} \int {d^4 p_E \over (2 \pi)^4}
  ~{1 \over  \pi R} ~ \sum_{n=-\infty}^{\infty} \left( \ln
\left[p_E^2 +  \left(\frac{n + \beta}{R}\right)^2 \right]
- \ln \left[p_E^2 +  \left(\frac{n}{R}\right)^2 \right] \right)  \cr
\noalign{\kern 5pt}
&&\hskip -.1cm
= \, {3\over 128 \pi^7 R^5} \sum_{n=1}^\infty {1\over n^5}
\, (1 - \cos 2\pi n \beta )   ~.
\label{v}
\end{eqnarray}
Here, $p_E$ is a four-dimensional Euclidean momentum.
Also the Wick rotation has been applied.  The quantities $\ep_0$ and $\Delta
\ep$ are divergent, whereas  $ v(\beta)$ is finite.

In gauge theory on the orbifold $S^1/Z_2$, there appears another
important representation, a $Z_2$ doublet.  A   $Z_2$ doublet field
$\phi = 
\begin{pmatrix}
\phi_1 \cr \phi_2
\end{pmatrix}$ satisfies
\beqn
&&\hskip -1cm
\phi(x,-y) = \tau_3 \phi(x,y) ~~, \cr
\noalign{\kern 5pt}
&&\hskip -1cm
\phi(x,y+2\pi R) = e^{-2\pi i\beta \tau_2} \, \phi(x,y)
= 
\begin{pmatrix}
 \cos 2\pi \beta & -\sin 2\pi \beta \cr
            \sin 2\pi \beta & ~~\cos 2\pi \beta 
\end{pmatrix} \,
       \phi(x,y) ~~.
\label{rep1}
\eeqn
Its expansion is given by
\beqn
\noalign{\kern 5pt}
&&\hskip -1cm
\begin{pmatrix}
\phi_1(x,y) \cr \phi_2(x,y) 
\end{pmatrix}
 =
{1\over \sqrt{\pi R}} \sum_{n=-\infty}^\infty
\phi_n(x) 
\begin{pmatrix}
\cos \mybig{(n+\beta) y\over R} \cr
                  \sin \mybig {(n+\beta) y\over R} 
\end{pmatrix}  ~~,
\label{expansion6}
\eeqn
or,  more concisely,  we write
\beeq
\begin{pmatrix}
\phi_1(x,y) \cr \phi_2(x,y) 
\end{pmatrix}
~=~ \bigg\{  \phi_n(x) ~;~ \beta \bigg\} ~~.
\label{expansion7}
\eneq
This type of BCs frequently appears in the  $SU(2)_R$ space in
supersymmetric theories.
As the mass for $\phi_n(x)$ is given by $[(n+\beta)/R]^2$,
$V_\eff$ for a $Z_2$ doublet bosonic field is given by
\beqn
V_\eff|_{Z_2 ~{\rm doublet}} &=&  -\frac{1}{2} \int {d^4 p_E \over (2
\pi)^4}
  ~{1 \over  \pi R} ~ \sum_{n=-\infty}^{\infty} \ln
\left[p_E^2 +  \left(\frac{n + \beta}{R}\right)^2 \right]   \cr
\noalign{\kern 5pt}
&=& \ep_0 + v(\beta) ,
\label{V2}
\end{eqnarray}
where $\ep_0$ and $v(\beta)$ are defined in (\ref{ep}) and (\ref{v}).

Let us apply the above result to non-supersymmetric $SU(N)$ gauge
theories.  To be definite,  the matter content in the bulk is assumed to
consist of
$n_s$ species of complex scalar fields  in the  fundamental representation,
$n_{fF}$ species of Dirac fermions  in the fundamental representation, and
$n_{fA}$ species of Dirac fermions  in the  2nd rank antisymmetric
representation.  We further suppose no additional flavor symmetry that
can generate
$Z_2$ doublets whose BCs are given in (\ref{rep1}).

The values $n^{(\eta_0 \eta_1)}_s$, $n^{(\eta_0 \eta_1)}_{fF}$ and
$n^{(\eta_0 \eta_1)}_{fA}$ are the numbers of scalar fields and Dirac
fermions, whose
BCs are given by
\beqn
&&\hskip -1cm
\phi(x, -y) = \eta_0 P_0 \phi(x, y) ~~, ~~
\phi(x, \pi R-y) = \eta_1 P_1 \phi(x, \pi R + y) \cr
\noalign{\kern 5pt}
&&\hskip -1cm
\psi_F(x, -y) = \eta_0 P_0 \gamma^5 \psi_F(x, y) ~~, ~~
\psi_F(x, \pi R-y) = \eta_1 P_1 \gamma^5 \psi_F (x, \pi R + y) ~~, \cr
\noalign{\kern 5pt}
&&\hskip -1cm
\psi_A(x, -y) = \eta_0 P_0 \gamma^5 \psi_A(x, y) P_0^t ~~, ~~
\psi_A(x, \pi R-y) = \eta_1 P_1 \gamma^5 \psi_A (x, \pi R + y) P_1^t~~,
\label{eta-s}
\eeqn
respectively.
Here, $\eta_0$ and $\eta_1$ take the value 1 or $-1$.
The sums of the numbers  $n^{(\eta_0 \eta_1)}_s$, $n^{(\eta_0
\eta_1)}_{fF}$ and $n^{(\eta_0 \eta_1)}_{fA}$
are denoted  $n_s$,  $n_{fF}$ and $n_{fA}$:
\beqn
n_s = \sum_{(\eta_0 \eta_1)} n^{(\eta_0 \eta_1)}_s ~~,~~
n_{fF} = \sum_{(\eta_0 \eta_1)} n^{(\eta_0 \eta_1)}_{fF} ~~,
n_{fA} = \sum_{(\eta_0 \eta_1)} n^{(\eta_0 \eta_1)}_{fA} ~~.
\eeqn

Let us introduce $N_{\rm rep}^{(P_0, P_1)}$ for each representation by
\beqn
&&\hskip -1cm
N_{Ad}^{(++)} = p^2 + q^2 + r^2 + s^2 - 1 \cr
&&\hskip -1cm
N_{Ad}^{(--)} = 2(ps + qr) \cr
&&\hskip -1cm
N_{Ad}^{(+-)} = 2(pq + rs) \cr
&&\hskip -1cm
N_{Ad}^{(-+)} = 2(pr + qs) \cr
\noalign{\kern 10pt}
&&\hskip -1cm
N_{F}^{(++)} = p \next
N_{F}^{(--)} = s \next
N_{F}^{(+-)} = q \next
N_{F}^{(-+)} = r \cr
\noalign{\kern 10pt}
&&\hskip -1cm
N_{A}^{(++)} = \onehalf \Big\{ p(p-1) + q(q-1) + r(r-1) + s(s-1) \Big\} \cr
&&\hskip -1cm
N_{A}^{(--)} = ps + qr \cr
&&\hskip -1cm
N_{A}^{(+-)} = pq + rs \cr
&&\hskip -1cm
N_{A}^{(-+)} = pr + qs
\label{counting1}
\eeqn
Under the $Z_2$ parity assignment (\ref{pqr}),
the quantities $N^{(P_0 P_1)}$ are found to be
\beqn
&&\hskip -1cm
N^{(P_0 P_1)} = N^{(P_0 P_1)}_g + N^{(P_0 P_1)}_s - N^{(P_0 P_1)}_f ~~,
\label{countN} \\
\noalign{\kern 10pt}
&&\hskip -1.cm
N^{(P_0 P_1)}_g = 2 N_{Ad}^{(P_0, P_1)} +  N_{Ad}^{(-P_0,-P_1)} ~~,
\label{N-g} \\
\noalign{\kern 5pt}
&&\hskip -1.cm
N^{(P_0, P_1)}_s =  \sum_{\eta_0, \eta_1}
  2 n^{(\eta_0, \eta_1)}_s  N_{F}^{(\eta_0 P_0, \eta_1 P_1)}   ~~,
\label{N-s} \\
\noalign{\kern 5pt}
&&\hskip -1.cm
N^{(P_0 P_1)}_{f} =  \sum_{\eta_0, \eta_1}
 2 n^{(\eta_0, \eta_1)}_{fF} (N_F^{(\eta_0 P_0, \eta_1 P_1)}
       + N_F^{(-\eta_0 P_0, - \eta_1 P_1)} )   \cr
\noalign{\kern 5pt}
&&\hskip 2cm
+ \sum_{\eta_0, \eta_1}
2 n^{(\eta_0, \eta_1)}_{fA} (N_A^{(\eta_0 P_0, \eta_1 P_1)}
   + N_A^{(- \eta_0 P_0, - \eta_1 P_1)} )
\label{N-f}
\eeqn
where  $N^{(P_0 P_1)}_g$, $N^{(P_0 P_1)}_s$ and $N^{(P_0 P_1)}_f$ are
the contributions from the gauge and FP ghost fields, complex scalar
fields and Dirac fermions.  The $A_y$ components of the gauge fields
are opposite in parity   to the four-dimensional
components $A_\mu$.   This leads to the expression (\ref{N-g}).
For fermions,  the  factor $\gamma^5$ in (\ref{eta-s}) implies that
 $(P_0, P_1)$ states are always accompanied by $(-P_0, -P_1)$ states.
This leads to (\ref{N-f}).

By use of (\ref{countN}), the formula for one-loop effective potential
at $A^0_M =0$ is given by
\beqn
&&\hskip -1cm
V_\eff = N_0 \ep_0 + N_{\Delta} \Delta \ep + N_v v(\onehalf) ~~, \cr
\noalign{\kern 10pt}
&&\hskip -1cm
N_0  = 3(N^2 - 1) + 2 n_s N - 4 n_{fF} N - 2 n_{fA}N(N-1)~~, \cr
&&\hskip -1cm
N_{\Delta} = (p-s)^2 + (q-r)^2 - 1
    + 2(n^{(++)}_s - n^{(--)}_s)(p-s) \cr
&&\hskip  1cm
 +  2(n^{(+-)}_s - n^{(-+)}_s)(q-r) ~~, \cr
&&\hskip -1cm
N_v =  (6 - 4n_{fA})(p+s)(q+r) \cr
&&\hskip 1cm
 + 2(n^{(++)}_s + n^{(--)}_s - 2 n^{(++)}_{fF} - 2
n^{(--)}_{fF})(q+r) \cr
&&\hskip 1cm
 + 2(n^{(+-)}_s + n^{(-+)}_s - 2
n^{(+-)}_{fF} - 2 n^{(+-)}_{fF})(p+s)   ~~.
\label{V-non}
\eeqn

At this stage, we recognize that $N_0$ is independent of the BCs,
and therefore independent  of $[p; q,r;s]$.  The $N_0$ term does not
distinguish BCs.  By contrast, $N_\Delta$ and $N_v$ do
depend on $[p; q,r;s]$.  There appears  a difference in the energy
density among theories in different equivalence classes.  It is tempting
to seek a theory with the lowest energy density which may be most
preferred, provided there exists dynamics connecting different equivalence
classes.  However, there arises fundamental ambiguity in the
$N_\Delta$ term:  $\Delta \ep$ is divergent.  There is no symmetry
principle that dictates  unique regularization.  One cannot compare the
energy densities in two theories in different equivalence classes.

In a previous paper \cite{HHHK}, we  evaluated $V_\eff$ in one
equivalence class as a function of the Wilson line phases.  The
value of $N_\Delta$ is the same in all theories in a given equivalence
class, and therefore there appeared no ambiguity there.  [See
(\ref{equ-rel}) and (\ref{V-non}).]  For instance, in the $SU(5)$ model, we
have the equivalence relations
\beeq
[p; q,r;s] = [2; 0,0; 3]
\sim [1;1,1;2] 
\sim [0;2,2;1] ~~.
\label{relate2}
\eneq
Suppose that all matter fields satisfy $\eta_0=\eta_1=1$ and set
$n_s^{(++)}=N_h$, $n_{fF}^{(++)}=N_f^5$, $n_{fA}^{(++)}=N_f^{10}$.
It then follows from (\ref{V-non}) for the effective potential
$V_\eff[p,q,r,s] \equiv V_\eff[A^0=0; p,q,r,s]$  that
\beqn
&&\hskip -1cm
V_\eff[ 1,1,1,2] - V_\eff[2,0,0,3]
= 4 (9 + N_h - 2 N_f^5 - 6 N_f^{10}) v(\onehalf) \cr
&&\hskip -1cm
V_\eff[0,2,2,1] - V_\eff[ 2,0,0,3]
= 8 (3 + N_h - 2 N_f^5 - 2 N_f^{10}) v(\onehalf) ~~.
\label{compare2}
\eeqn
These agree with the result in Ref.\ 24).\footnote{$v(\beta)$ is
related to $f_5(\beta)$ in Ref.\ 24) by
$v(\beta) = {1\over 2} C [f_5(0) - f_5(2 \beta)]$ and $C=3/64 \pi^7 R^5$.
There was a factor 2 error in the normalization of the effective potential
in Ref.\ 24).}
When $N_f^5 = N_f^{10}=0$, for instance, the theory with
$[p;q,r;s] = [2;0,0;3]$, which has the
$SU(3) \times SU(2) \times U(1)$ symmetry,  has the lowest energy density.

However, if one tries to compare theories in different equivalence
classes,  the ambiguity in the $N_\Delta$ term cannot be
avoided.  This ambiguity naturally disappears in supersymmetric theories, as
we  spell out below.

\section{Supersymmetric gauge theory}

In this section, we derive generic formulas for the one-loop effective
potential at vanishing Wilson line phases in supersymmetric (SUSY)
$SU(N)$ gauge theories and compare the vacuum energy density  in  
the theories belonging to various equivalence classes of the BCs.

If the theory has unbroken supersymmetry, then the effective potential for
Wilson lines remains flat due to the cancellation among contributions
from bosonic fields and fermionic fields.
A nontrivial dependence of $V_\eff$ appears if SUSY is softly broken
as the nature demands.
There is a natural way to introduce soft SUSY breaking on an orbifold.
$N=1$ SUSY in five-dimensional space-time corresponds to $N=2$ SUSY in
four-dimensional spacetime.
A five-dimensional gauge multiplet
${\cal{V}}=(A_M, \lambda, \lambda', \sigma) \equiv
(A_M, \lambda^1_L, \lambda^2_L, \sigma)$ is decomposed into a vector
superfield
$V=(A_\mu, \lambda)$
and a chiral superfield $\Sigma = (\Phi \equiv \sigma + iA_y, \lambda')$
in four dimensions.
Similarly, a hypermultiplet
${\cal H}=(h, h^c{}^\dagger, \tilde{h}, \tilde{h}^c{}^\dagger)
\equiv (h_1, h_2, \tilde{h}_L, \tilde{h}_R )$
is decomposed into two chiral superfields in four dimensions as
$H=(h, \tilde{h})$
and $H^c=(h^c, \tilde{h}^c)$
where $H$
and $H^c$
undergo conjugated transformation under $SU(N)$.
After a translation along a non-contractible loop, these fields and
their superpartners may have different twist, depending on their
$SU(2)_R$ charges.\cite{SSorbifolds,Nomura1}
This is called the Scherk-Schwarz breaking mechanism.\cite{SS}
We  adopt this mechanism for the SUSY breaking,
which makes the evaluation of the effective potential easy.

The theory can contain several kinds of hypermultiplets in
various kinds of representations in the bulk, some of which play the role
of the Higgs multiplets or of the  quark-lepton multiplets.
Further there may exist $N=1$ supermultiplets on the boundary 
branes.\footnote{ It is known that anomalies may arise at the boundaries
with chiral fermions.\cite{anomaly}
These anomalies must be cancelled in the four-dimensional effective theory,
for instance, by such local counter terms as the
Chern-Simons term.\cite{anomaly,KKL}
We assume that the four-dimensional effective theory is anomaly free.}
Here we write down the bulk part of the typical Lagrangian density
${\cal L}_{\rm bulk}$
for ${\cal V}$ and ${\cal H}$ to discuss their BCs:
\begin{eqnarray}
{\cal L}_{\rm bulk}
&=& \frac{1}{g^2}\left(-{1 \over 2} {\rm Tr} F_{MN}^2
 + {\rm Tr} |D_M \Phi|^2 + {\rm Tr} (i\bar{\lambda}^i \gamma^M D_M
\lambda^i)
   - {\rm Tr} (\bar{\lambda}^i [\Phi, \lambda^i])\right) \cr
\noalign{\kern 10pt}
&& + |D_M h_i|^2  + \bar{\tilde H} (i\gamma^M D_M - \Phi) {\tilde H}
 - (i \sqrt{2} h_i^{\dagger} \bar{\lambda}^i \bar{\tilde H} + \mbox{h.c.})
\cr
\noalign{\kern 10pt}
&&  - h_i^{\dagger} \Phi^2 h_i - \frac{g^2}{2} \sum_{m, \alpha}
 \left(h_i^{\dagger} (\tau^m)_{ij} T_h^{\alpha} h_j \right)^2  ~~.
\label{bulkL}
\end{eqnarray}
Here the quantities $\lambda^i$ are the symplectic Majorana spinors defined
in Ref.\ 31):
$\lambda^j = \mybig 
\begin{pmatrix}
\lambda^j_L \cr
  i  \ep^{jk} \sigma^2 {\lambda^k_L}^* 
\end{pmatrix}$.
$\tilde{H}$ is  a Dirac spinor,
$\tilde{H} = \mybig 
\begin{pmatrix}
\tilde{h}_L \cr \tilde{h}_R
\end{pmatrix}$,
and $T_h^{\alpha}$'s are representation matrices of $SU(N)$ gauge
generators for $h$.

The requirement that the Lagrangian density  be single valued allows
the following nontrivial BCs.
For the gauge multiplet, we have
\beqn
&&\hskip -1cm
\begin{pmatrix}
V \cr \Sigma
\end{pmatrix}
 (x, -y)
= P_0 ~
\begin{pmatrix}
 V \cr -\Sigma
\end{pmatrix}
 (x, y) ~ P_0^\dagger \cr
\noalign{\kern 10pt}
&&\hskip -1cm
A_\mu(x, \pi R-y)= P_1 ~ A_\mu(x, \pi R+y) ~ P_1^\dagger ~~, \cr
\noalign{\kern 3pt}
&&\hskip -1cm
A_y(x, \pi R-y)= -P_1 ~ A_y(x, \pi R+y) ~ P_1^\dagger ~~, \cr
\noalign{\kern 5pt}
&&\hskip -1cm
\left(
\begin{array}{c}
\lambda \\
\lambda'
\end{array}
\right) (x, \pi R-y)
= e^{-2\pi i \beta \tau_2} ~
P_1 ~\left(
\begin{array}{c}
\lambda \\
-\lambda'
\end{array}
\right) (x, \pi R+y) ~P_1^\dagger ~~,  \cr
\noalign{\kern 5pt}
&&\hskip -1cm
\sigma(x,\pi R-y)= -P_1~ \sigma(x,\pi R+y) ~ P_1^\dagger ~~,\cr
\noalign{\kern 10pt}
&&\hskip -1cm
A_M(x, y+2 \pi R)= U ~ A_M(x, y) ~ U^\dagger ~~, \cr
\noalign{\kern 3pt}
&&\hskip -1cm
\left(
\begin{array}{c}
\lambda \\
\lambda'
\end{array}
\right) (x, y+2\pi R)
= e^{-2\pi i \beta \tau_2} ~
U ~\left(
\begin{array}{c}
\lambda \\
\lambda'
\end{array}
\right) (x,y) ~U^\dagger ~~,  \cr
\noalign{\kern 5pt}
&&\hskip -1cm
\sigma(x,y+2\pi R)= U~ \sigma(x,y) ~ U^\dagger ~~.
\label{SBC1}
\end{eqnarray}
For a hypermultiplet ${\cal H}$, the BCs are given by
\beqn
&&\hskip -1cm
\begin{pmatrix}
h \cr h^c{}^\dagger
\end{pmatrix}
 (x, -y)
= \eta_0 T_{\cal H}[P_0] ~
\begin{pmatrix}
h \cr -h^c{}^\dagger
\end{pmatrix}
  (x, y) ~~, \cr
&&\hskip -1cm
\begin{pmatrix}
h \cr h^c{}^\dagger
\end{pmatrix}
(x, \pi R-y)
= e^{-2\pi i \beta \tau_2} ~ \eta_1 T_{\cal H}[P_1]
  ~
\begin{pmatrix}
h \cr -h^c{}^\dagger
\end{pmatrix}  (x, \pi R + y) ~~, \cr
&&\hskip -1cm
\begin{pmatrix}
h \cr h^c{}^\dagger
\end{pmatrix}
 (x, y + 2\pi R)
= e^{-2\pi i \beta \tau_2} ~ \eta_0 \eta_1 T_{\cal H}[U]
 ~
\begin{pmatrix}
h \cr h^c{}^\dagger
\end{pmatrix}
  (x, y) ~~, \cr
\noalign{\kern 10pt}
&&\hskip -1cm
\begin{pmatrix}
\tilde{h} \cr \tilde{h}^c{}^\dagger
\end{pmatrix}
 (x, -y)
= \eta_0 T_{\cal H}[P_0]
~
\begin{pmatrix}
\tilde{h} \cr -\tilde{h}^c{}^\dagger
\end{pmatrix}
  (x, y) ~~, \cr
&&\hskip -1cm
\begin{pmatrix}
\tilde{h} \cr \tilde{h}^c{}^\dagger
\end{pmatrix}
 (x, \pi R-y)
= \eta_1 T_{\cal H}[P_1]
~
\begin{pmatrix}
\tilde{h} \cr -\tilde{h}^c{}^\dagger
\end{pmatrix}
  (x, \pi R + y) ~~, \cr
&&\hskip -1cm
\begin{pmatrix}
\tilde{h} \cr \tilde{h}^c{}^\dagger
\end{pmatrix}
 (x, y + 2\pi R)
= \eta_0 \eta_1 T_{\cal H}[U]
~
\begin{pmatrix}
\tilde{h} \cr \tilde{h}^c{}^\dagger
\end{pmatrix}
  (x, y) ~~.
\label{SBC2}
\end{eqnarray}
Here $T_{\cal H}[P_0] h$ represents $P_0 h$ or $P_0 h P_0^\dagger$ for
$h$ in the fundamental or adjoint representation, respectively.
A hypermultiplet with  parity $(\eta_0, \eta_1)$ gives the 
same contribution to the vacuum energy density in the bulk as 
a hypermultiplet with  parity $(-\eta_0, -\eta_1)$.

With a nonvanishing $\beta$, there appear soft SUSY breaking mass terms
for gauginos $\lambda^i$ and scalar fields $h_i$
in the four-dimensional theory.\footnote{
$\beta$ should be of order $10^{-14}$, on the basis of phenomenological
consideration, for the soft SUSY breaking masses to be $O(1)$ TeV if 
$1/R \sim 10^{16}\,$GeV.} From the above
BCs (\ref{SBC1}) and (\ref{SBC2}),
mode expansions of each field are obtained.
Let $(P_0,P_1)_\lambda$ be the parity assignment of each component of
$\lambda(x,y)$ defined by $P_0 \lambda P_0$ and $P_1 \lambda P_1$.
Depending on $(P_0,P_1)_\lambda$,
the mode expansion for each component of gauginos is given by
\beqn
(P_0,P_1)_\lambda = (+1,+1) : \quad
\begin{pmatrix}
 \lambda (x, y)\cr {\lambda'} (x, y) 
\end{pmatrix}
&=& \bigg\{  \lambda_n(x) ~;~ \beta \bigg\}  ~~, \cr
\noalign{\kern 10pt}
  (-1,-1) : \quad
\begin{pmatrix}
 \lambda' (x, y)\cr {\lambda} (x, y) 
\end{pmatrix}
&=& \bigg\{  \lambda_n(x) ~;~ - \beta \bigg\}  ~~, \cr
\noalign{\kern 10pt}
(+1,-1) : \quad
\begin{pmatrix}
 \lambda (x, y)\cr {\lambda'} (x, y) 
\end{pmatrix}
&=& \bigg\{  \lambda_n(x) ~;~ \beta + \onehalf \bigg\}  ~~, \cr
\noalign{\kern 10pt}
(-1,+1) : \quad
\begin{pmatrix}
 \lambda' (x, y)\cr {\lambda} (x, y)
\end{pmatrix}
&=& \bigg\{  \lambda_n(x) ~;~ -\beta - \onehalf \bigg\}  ~~.
\label{SUSYexpansion1}
\eeqn
Hence the contribution to  $V_\eff$ from the gauginos is given by
\beqn
&&\hskip -1cm
V_\eff|_{\rm gauginos} =
- 4 \bigg\{ N^{(++)}_{Ad} (\ep_0
     + v(\beta)) + N^{(--)}_{Ad}  (\ep_0 + v(-\beta))  \cr
\noalign{\kern 5pt}
&&\hskip 2.0cm
 + N^{(+-)}_{Ad} (\ep_0 + v(\beta+ \onehalf))
+ N^{(-+)}_{Ad} (\ep_0 + v(-\beta- \onehalf)) \bigg\} ~~,
\label{Vgaugino}
\eeqn
where the factor 4 comes from the number of the degrees of freedom for each
Majorana fermion.  The values $N^{(P_0 P_1)}_{Ad}$ are defined in
(\ref{counting1}).

The boson part of a gauge multiplet contains an additional scalar field
$\sigma$,
which has the same parity assignment as $A_y$.  Hence the contributions
from the boson part are
\beqn
&&\hskip -1cm
V_\eff|_{\rm gauge, ghost, \sigma} =
(2 N^{(++)}_{Ad} + 2 N^{(--)}_{Ad} ) (\ep_0 + \Delta \ep)
+ (2 N^{(--)}_{Ad} + 2 N^{(++)}_{Ad} ) (\ep_0 - \Delta \ep) \cr
\noalign{\kern 10pt}
&&\hskip 1cm
+ (2 N^{(+-)}_{Ad} + 2 N^{(-+)}_{Ad} ) (\ep_0 + v(\onehalf))
+ (2 N^{(-+)}_{Ad} + 2 N^{(+-)}_{Ad} ) (\ep_0 + v(\onehalf))~~.
\label{Vgauge}
\eeqn
Notice that the $\Delta \ep$ terms cancel among the contributions from
the bosons in the supersymmetric theory.

The gauge multiplet part of $V_\eff$ is obtained by adding  (\ref{Vgaugino})
and (\ref{Vgauge}):
\beqn
&&\hskip -1cm
V_\eff|_{\rm gauge~multiplet}
= - 4 (N^2 - 1) v(\beta)
 + 4 (N^{(+-)}_{Ad} + N^{(-+)}_{Ad}) w(\beta) \cr
\noalign{\kern 10pt}
&&\hskip 1.9cm
=- 4 (N^2 - 1) v(\beta)   + 4 (p+s)(q+r) w(\beta)  ~~, \cr
\noalign{\kern 10pt}
&&\hskip .9cm
w(\beta) \equiv v(\onehalf) + v(\beta) - v(\beta+\onehalf) \ge 0 ~~,
\label{Vgaugemultiplet}
\eeqn
where  the relation $v(-\beta) = v(\beta)$ has been  used.
$V_\eff|_{\rm gauge~multiplet}$
takes its minimum value, $- 4 (N^2 - 1) v(\beta)$,
at   $p = s = 0$, $q + r =N$ or $p + s = N$, $q = r = 0$.
Each case corresponds to
\beqn
&&\hskip -1cm
\diag P_0 = (\overbrace{+1, \cdots \cdots, +1 ,
        -1, \cdots \cdots, -1}^N) ~, \nonumber \\
&&\hskip -1cm
\diag P_1 = ( \underbrace{-1, \cdots \cdots, -1}_q,
\underbrace{+1, \cdots \cdots, +1}_{r}) ~~
\label{p=s=0}
\eeqn
or
\beqn
&&\hskip -1cm
\diag P_0 = (\overbrace{+1, \cdots \cdots, +1 ,
        -1, \cdots \cdots, -1}^N) ~, \nonumber \\
&&\hskip -1cm
\diag P_1 = ( \underbrace{+1, \cdots \cdots, +1}_p,
\underbrace{-1, \cdots \cdots, -1}_{s}) ~,
\label{q=r=0}
\eeqn
respectively.
The gauge symmetry is broken  to $SU(q) \times SU(N-q) \times U(1)$
or $SU(p) \times SU(N-p) \times U(1)$, respectively.

It is interesting to examine what types of breaking patterns are 
induced  by the introduction of hypermultiplets.
The mode expansion for the hypermultiplet
${\cal H} = (h, h^{c\dagger}, \tilde h, \tilde h {}^{c\dagger})$ is found
in a similar manner.
Let $(P_0, P_1)_h = (a,b)_h$ be the parity assignment of each component of
$h(x,y)$ defined by $\eta_0 (T_{\cal H}[P_0] h)^j = a h^j$ and
 $\eta_1 (T_{\cal H}[P_1] h)^j = b h^j$.
Then, depending on $(P_0, P_1)_h$, one finds
\beqn
(P_0,  P_1)_h = (+1,+1) : &&
\begin{pmatrix}
 h \cr h^{c\dagger} 
\end{pmatrix}
= \bigg\{  h_n(x) ~;~ \beta \bigg\}  ~~,~~
\cr
\noalign{\kern 10pt}
 (-1,-1) : &&
\begin{pmatrix}
 h^{c\dagger} \cr  h 
\end{pmatrix}
= \bigg\{  h_n(x) ~;~ - \beta \bigg\}  ~~,~~ \cr
\noalign{\kern 10pt}
 (+1,-1) : &&
\begin{pmatrix}
 h \cr h^{c\dagger} 
\end{pmatrix}
= \bigg\{  h_n(x) ~;~ \beta + \onehalf \bigg\}  ~~,~~
\cr
\noalign{\kern 10pt}
 (-1,+1) : &&
\begin{pmatrix}
 h^{c\dagger} \cr  h 
\end{pmatrix}
= \bigg\{  h_n(x) ~;~ - \beta -\onehalf \bigg\}  ~~.
\label{SUSYexpansion2}
\eeqn
For their fermionic superpartners, the mode expansions are obtained by
setting
$\beta = 0$ in  (\ref{SUSYexpansion2}).
Hence the contribution to  $V_\eff$ from ${\cal H}$ is given by
\beqn
&&\hskip -1cm
V_\eff|_{\cal H} =
  4 \Big( N_h^{(++)} + N_h^{(--)} \Big) v(\beta)
  + 4 \Big( N_h^{(+-)} + N_h^{(-+)}\Big)
    \Big( v(\beta+ \onehalf) - v(\onehalf) \Big)  \cr
\noalign{\kern 10pt}
&&\hskip 0.1cm
= 4 N_h^{\rm total} v(\beta) - 4 \Big( N_h^{(+-)} + N_h^{(-+)}\Big)
 w(\beta) ~~.
\label{Vhyper}
\eeqn
Here $N_h^{(P_0, P_1)}$ is the number of
components of $h(x,y)$ with  parity $(P_0, P_1)$.
The value $N_h^{\rm total} = \sum_{a,b} N_h^{(ab)}$ is independent of
$(p,q,r,s)$. We note that
\beeq
N_h^{(P_0, P_1)} =
\sum_{{\rm rep}=Ad, F, A}
n_{\rm rep}^{(\eta_0, \eta_1)} N_{\rm rep}^{(\eta_0 P_0, \eta_1 P_1)}
\label{counting2}
\eneq
where the quantities $N_{\rm rep}^{(a,b)}$ are given in (\ref{counting1}).

Suppose that the matter content in the bulk is given by
$n_{Ad}^{(\pm)}$, $n_{F}^{(\pm)}$ and $n_{A}^{(\pm)}$  species of
hypermultiplets  with $\eta_0 \eta_1 = \pm 1 $ in the adjoint, fundamental,
and 2nd rank antisymmetric representations, respectively.
(Note that $n_{\rm rep}^{(+)} = n_{\rm rep}^{(++)} + n_{\rm rep}^{(--)}$ and
 $n_{\rm rep}^{(-)} = n_{\rm rep}^{(+-)} + n_{\rm rep}^{(-+)}$.)
It is straightforward to extend our analysis including hypermultiplets in
bigger representations.
The total effective potential is given by
\beqn
&&\hskip -1cm
V_\eff
= 4 \Bigg\{ (-1 + n_{Ad}^{(+)}  + n_{Ad}^{(-)}) (N^2 - 1)
  + (n_{F}^{(+)} + n_{F}^{(-)}) N
  + (n_{A}^{(+)} + n_{A}^{(-)})  \frac{N(N-1)}{2} \Bigg\} v(\beta) \cr
\noalign{\kern 5pt}
&&\hskip -0cm
 + 4 \Bigg\{ (p+s)(q+r) (2- 2n_{Ad}^{(+)} + 2n_{Ad}^{(-)}
   - n_{A}^{(+)}  + n_{A}^{(-)} )  \cr
\noalign{\kern 5pt}
&&\hskip +2cm
   - n_{Ad}^{(-)} (N^2 -1) - n_{A}^{(-)} \frac{N(N-1)}{2}
- n_{F}^{(+)} (q+r)  -  n_{F}^{(-)} (p+s) \Bigg\} w(\beta)  \cr
\noalign{\kern 10pt}
&&\hskip -.5cm
= ~ (p,q,r,s\hbox{-independent terms}) + 4 w(\beta) ~ h(q+r)~~.
\label{effVtotal1}
\eeqn
Here the function $h(x)$ ($0 \le x=q+r = N-(p+s) \le N$) is defined by
\beqn
&&\hskip -1cm
h(x) = a x (N-x) - b x  ~~, \cr
\noalign{\kern 10pt}
&&\hskip -1.0cm
\begin{cases}
a =  2- 2n_{Ad}^{(+)} + 2n_{Ad}^{(-)} - n_{A}^{(+)}  + n_{A}^{(-)} ~~, \cr
\noalign{\kern 3pt}
b =  n_{F}^{(+)} - n_{F}^{(-)}  ~~. 
\end{cases}
\label{effVtotal2}
\eeqn
As $w(\beta) >0$ for nonintegral $\beta$,
the minimum of the energy density is given by that of $h(x)$, which is
determined by $a$ and $b$.

Let us classify various cases.

\noindent
(i) The case with $a=0$

In this case the BCs which give the minimum energy density are
\beeq
\begin{cases}
n_{F}^{(+)} >  n_{F}^{(-)}  &\Rightarrow \quad q+r=N~,\cr
\noalign{\kern 3pt}
n_{F}^{(+)} =  n_{F}^{(-)}  &\Rightarrow \quad \hbox{completely
degenerate}~,\cr
\noalign{\kern 3pt}
n_{F}^{(+)} <  n_{F}^{(-)}  &\Rightarrow \quad q+r=0  ~~.
\end{cases}
\label{effVmin1}
\eneq

\noindent
(ii) The case with $a > 0$

In this case, $h(x)$ is minimized either at $x=0$ or at $x=N$.
As $h(N) - h(0) = -N b $, we have
\beeq
\begin{cases}
n_{F}^{(+)} >  n_{F}^{(-)}  &\Rightarrow \quad q+r=N ~,\cr
\noalign{\kern 3pt}
n_{F}^{(+)} =  n_{F}^{(-)}  &\Rightarrow \quad q+r = 0, N ~, \cr
\noalign{\kern 3pt}
n_{F}^{(+)} <  n_{F}^{(-)}  &\Rightarrow \quad q+r=0  ~~.
\end{cases}
\label{effVmin2}
\eneq

\noindent
(iii) The case with $a < 0$

In this case, we have
\beeq
h(x) = |a|  \Big\{ (x - x_0)^2  - x_0^2 \Big\} 
\next x_0 = {N\over 2}  + {b  \over 2 |a|} ~~,
\label{hmin}
\eneq
so that $h(q+r)$ can have a minimum at $q+r$ between 1 and $N-1$.
Let $[x_0]_{\rm nearest}$ be the integer nearest to $x_0$.  Then
\beeq
\begin{cases}
x_0 \le 0        &\Rightarrow \quad q+r=0 ~,\cr
0 < x_0 < N      &\Rightarrow \quad q+r=  [x_0]_{\rm nearest} ~,\cr
x_0 \ge N        &\Rightarrow \quad q+r=N~~.
\end{cases}
\label{effVmin3}
\eneq

Let us examine a couple of examples.
In the $SU(5)$ GUT proposed in
Ref.\  13), only the gauge multiplet and the fundamental Higgs
multiplets exist in the bulk, whereas the quark and lepton multiplets are
confined on one of the boundary branes.  The parity of the Higgs
hypermultiplets is assigned such that $n_F^{(++)} = n_F^{(--)} = 1$.
This corresponds to the case with $n_F^{(+)}=2$ and
 $n_F^{(-)}=n_A^{(\pm)}=n_{Ad}^{(\pm)}=0$.   In one of the models
considered in Ref.\ 17),
the quarks and leptons also reside in the bulk.
The three generations of quarks and leptons add 
$n_F^{(++)} = n_F^{(+-)} = n_A^{(++)} = n_A^{(+-)} = 3$. 
In all, $n_F^{(+)}=5$, $n_F^{(-)}=3$, $n_A^{(\pm)}=3$  and
 $n_{Ad}^{(\pm)}=0$. The same matter content was examined in Ref.\
24).
In all of those models, $a=b=2$, and thus the equivalence classes
with $q+r=N$ have the lowest energy density.
In Refs.\ 13) and 17), the BC
$[p;q,r;s]=[2;3,0;0]$ has been adopted to reproduce MSSM at low energies.
In Ref.\ 24), the BC $[p;q,r;s]=[2;0,0;3]$ is
examined.  The result in the present paper shows that with the given matter
content,
the equivalence classes to which these BCs belong are not
those with the lowest energy density.  They may not be selected, provided
that
there exist dynamics connecting different equivalence classes.

This, however, does not preclude the possibility of having these BCs.
As pointed out in Refs.\ 13) and 14),
$[p;q,r;s]=[2;3,0;0]$ has the nice feature of reproducing MSSM at low
energies with the natural triplet-doublet splitting.   
To obtain exactly three families of matter chiral multiplets and two weak Higgs 
chiral multiplets as zero modes (namely, massless particles or light 
particles with masses of O($\beta/R$) in four dimensions) 
 of hypermultiplets in the bulk
and of chiral multiplets on the brane, the  relations 
\beqn
&~& n_{5}^{(++)} + n_{\overline{5}}^{(--)} = 1 ~~, \cr
&~& n_{\overline{5}}^{(++)} + n_{5}^{(--)} 
    + n_{\overline{5}}^{({\rm Brane})} = 4 ~~, \cr
&~& n_{\overline{5}}^{(+-)} + n_{5}^{(-+)} 
    + n_{\overline{5}}^{({\rm Brane})} = 3 ~~, \cr
&~& n_{10}^{(++)} + n_{\overline{10}}^{(--)} 
    + n_{10}^{({\rm Brane})} = 3 ~~, \cr
&~& n_{10}^{(+-)} + n_{\overline{10}}^{(-+)} 
    + n_{10}^{({\rm Brane})} = 3 ~~, \cr
&~& n_{24}^{(+)} = n_{24}^{(-)} = 0 ~~,
\label{MSSMcondition}
\eeqn
must hold, as can be inferred from  Table I.
Here $n_{\overline{5}}^{({\rm Brane})}$ and $n_{10}^{({\rm Brane})}$  are the
numbers of chiral multiplets on the brane whose representations are
$\overline{\bf 5}$ and ${\bf 10}$, respectively.
\begin{table}[ht]
\begin{center}
\begin{tabular}{c|cccc}
 & $(\eta_0, \eta_1)=(++)$ & $(--)$ & $(+-)$ & $(-+)$ \\ \hline\hline
${\bf 5}^{(\eta_0, \eta_1)}$ & $({\bf 1}, {\bf 2})_{1/2}$
 & $({\bf 1}, {\bf 2})_{-1/2}$ 
& $({\bf 3}, {\bf 1})_{-1/3}$ & $({\overline{\bf 3}}, {\bf 1})_{1/3}$ \\ \hline
${\overline{\bf 5}}^{(\eta_0, \eta_1)}$ & $({\bf 1}, {\bf 2})_{-1/2}$
& $({\bf 1}, {\bf 2})_{1/2}$ 
& $({\overline{\bf 3}}, {\bf 1})_{1/3}$ & $({\bf 3}, {\bf 1})_{-1/3}$ \\ \hline
${\bf 10}^{(\eta_0, \eta_1)}$
& $({\overline{\bf 3}}, {\bf 1})_{-2/3} + ({\bf 1}, {\bf 1})_{1}$ 
& $({\bf 3}, {\bf 1})_{2/3} + ({\bf 1}, {\bf 1})_{-1}$ 
& $({\bf 3}, {\bf 2})_{1/6}$ & $({\overline{\bf 3}}, {\bf 2})_{-1/6}$ \\ \hline
$\overline{\bf 10}^{(\eta_0, \eta_1)}$ 
& $({\bf 3}, {\bf 1})_{2/3} + ({\bf 1}, {\bf 1})_{-1}$ 
& $({\overline{\bf 3}}, {\bf 1})_{-2/3} + ({\bf 1}, {\bf 1})_{1}$ 
& $({\overline{\bf 3}}, {\bf 2})_{-1/6}$ & $({\bf 3}, {\bf 2})_{1/6}$ \\ \hline
${\bf 24}^{(\eta_0, \eta_1)}$ & $({\bf 8}, {\bf 1})_{0} + ({\bf 3}, {\bf 1})_{0}$ 
& $({\bf 8}, {\bf 1})_{0} + ({\bf 3}, {\bf 1})_{0}$ 
& $({\bf 3}, {\bf 2})_{-5/6}$ 
& $({\bf 3}, {\bf 2})_{-5/6}$ \\ 
& $+ ({\bf 1}, {\bf 1})_{0}$ & $+ ({\bf 1}, {\bf 1})_{0}$ 
& $+ ({\overline{\bf 3}}, {\bf 2})_{5/6}$ 
& $+ ({\overline{\bf 3}}, {\bf 2})_{5/6}$ \\ 
\end{tabular}
\end{center}
\caption[table-SU(5)multiplets]{Standard model gauge quantum numbers of
zero modes in $SU(5)$ hypermultiplets with ($\eta_0$, $\eta_1$).}
\label{table-SU(5)multiplets}%
\end{table}
In order for the class $[p;q,r;s]=[2;3,0;0]$ to have the lowest energy density,
we need $a <0$ and $[x_0]_{\rm nearest} = 3$.
In other words we need
\beqn
2\Big\{ 2n_{Ad}^{(+)} - 2n_{Ad}^{(-)}
      + n_{A}^{(+)} - n_{A}^{(-)} - 2 \Big\}
\geq n_{F}^{(+)} - n_{F}^{(-)} \geq 0 ~~.
\label{ineq2}
\eeqn

The set of relations in (\ref{MSSMcondition}) is incompatible with 
 the first inequality in (\ref{ineq2}).
In order to reconcile these relations, we need an extension of the model
with extra  hypermultiplets (and brane fields).
Then there appear additional zero modes, which might threaten the stability
of protons and the successful gauge coupling unification based on MSSM.
If these zero modes acquire large masses, say, through  coupling 
with extra singlets on the brane, the phenomenological 
disaster can be avoided.

Let us give an example with $a <0$ and $[x_0]_{\rm nearest} = 3$.
It is realized with $a = -2$ and $b = 2$, that is,
$(n_{Ad}^{(+)} - n_{Ad}^{(-)}, n_{A}^{(+)} - n_{A}^{(-)}) = (0, 4)$, 
$(1, 2)$, $(2, 0)$ and $n_{F}^{(+)} - n_{F}^{(-)} = 2$.
The latter equality holds in (\ref{MSSMcondition}).
Consider $(n_{Ad}^{(-)}, n_{A}^{(-)}) = (0, 0)$ for  simplicity.
To have $(n_{Ad}^{(+)}, n_{A}^{(+)}) = (0, 4)$,
we need to pick  two extra sets of   hypermultiplet pairs in 
$({\bf 10}^{(++)}, {\bf 10}^{(--)})$, 
$({\bf 10}^{(++)},  {\overline{\bf 10}}^{(++)})$,
$({\overline{\bf 10}}^{(++)}, {\overline{\bf 10}}^{(--)})$
or $({\bf 10}^{(--)}, {\overline{\bf 10}}^{(--)})$.
There appear two pairs of zero modes with representations  
$({\overline{\bf 3}}, {\bf 1})_{-2/3} + ({\bf 1}, {\bf 1})_{1}$ 
and $({\bf 3}, {\bf 1})_{2/3} + ({\bf 1}, {\bf 1})_{-1}$
with the standard model gauge group.
They can form SUSY mass terms through the interactions
$({\overline{\bf 3}}, {\bf 1})_{-2/3} \cdot ({\bf 3}, {\bf 1})_{2/3}
  \cdot ({\rm singlet})$ and 
$({\bf 1}, {\bf 1})_{1} \cdot ({\bf 1}, {\bf 1})_{-1} \cdot ({\rm singlet})$
and decouple from the low energy theory if the magnitudes of the
VEVs of singlets on the brane are sufficiently large.
In other cases with  $(n_{Ad}^{(+)}, n_{A}^{(+)}) = (1, 2)$ and 
$(n_{Ad}^{(+)}, n_{A}^{(+)}) = (2, 0)$ as well,
phenomenologically interesting low energy theories can be derived  
in a similar manner.  A careful analysis  is necessary to check whether or not
this  scenario is realistic.

Another possibility is to have $[p;q,r;s]=[2;0,0;3]$ as the preferred
equivalence class.  For this end we need
$a \ge 0$ and $n_{F}^{(+)} \le n_{F}^{(-)}$.  This is realized if,
for instance, $n_F^{(\pm)}=5$,  $n_A^{(\pm)}=3$  and  $n_{Ad}^{(\pm)}=0$.
In this scenario, however, there appear additional light particles
 which may threaten the stability of
protons without further implementing such symmetry as $U(1)_R$ to forbid it
and also may ruin the gauge coupling unification in MSSM.

Further, irrespective of the matter content in the bulk, there 
remains the degeneracy.  The effective potential (\ref{effVtotal1}) is
a function of $q+r$ only.  It does not select  unique $[p;q,r;s]$.
We need to find a  mechanism  to lift the degeneracy.

\section{Cosmological evolution}

In the present paper we have evaluated
the vacuum energy density in each equivalence class of BCs
in SUSY $SU(N)$ gauge theory on the orbifold
$S^1/Z_2$.   With the soft SUSY breaking there arise
energy differences among different vacua.

An imminent question of great concern is how the BCs
are selected or determined.  Although it is natural to expect that
the BCs yielding the lowest energy density would be
selected, the problem is not so simple, as the mechanism for
transitions among different BCs is not well understood.

Here we have to distinguish two kinds of transitions.  In each
equivalence class there are, in general, infinitely many theories with
different BCs.  There are Wilson line phases whose
dynamics yield and guarantee the same physics in the equivalence
class.  The effective potential for the Wilson line phases has, in
general, more than one minimum.  These minima are separated by barriers
whose heights are about $\beta^2/R^4$ in the four-dimensional energy density
for $|\beta| \ll 1$.  In
the GUT picture,
$M_\GUT \sim 1/R$ and $M_\SUSY \sim \beta/R$.  The energy scale
characterizing the barrier is $V^{(1)}_\B \sim \sqrt{\beta}/R \sim
\sqrt{M_\GUT M_\SUSY}$. Transitions among different minima occur either
thermally or quantum mechanically.  The quantum tunneling transition
rate at zero temperature, however, is negligibly small.\cite{HHHK}

Nothing definite can be said about transitions among different
equivalence classes without an understanding of the dynamics connecting
these different equivalence classes.  We are supposing that
there exist such dynamics. This must certainly be true if the
structure of the spacetime is determined dynamically as in 
string theory.  One needs to know  the  height,  $V^{(2)}_\B$,
of the effective  barrier separating the different equivalence classes.
It would be below $M_\GUT$ where the spacetime structure $S^1/Z_2$ is
selected. As the typical energy difference among different
equivalence classes is $\beta^2/R^4$ for $|\beta| \ll 1$, it should be
above
$V^{(1)}_\B$. Hence we have
\beeq
V^{(1)}_\B  \sim \sqrt{M_\GUT M_\SUSY}
\le V^{(2)}_\B \le M_\GUT ~~.
\label{barrier1}
\eneq
The quantum tunneling rate at zero temperature from one equivalence
class to another is probably negligibly small.

To understand how the low energy symmetry is determined, one needs to
trace the cosmological evolution of the universe.  At the very early
stage around the scale $M_\GUT \sim 1/R$, it is suspected that the
effective five-dimensional orbifold $M^4 \times (S^1/Z_2)$ emerges.
The universe then can be either cold or hot.

If the universe is cold, the selection of the equivalence class is due
solely  to the dynamics connecting different classes.  Without
understanding the dynamics, nothing can be said for sure about the
selection.

If the universe is hot with temperature $T \sim M_\GUT$, there is an
ample amount of thermal transitions among theories with different
BCs in various equivalence classes.  When the temperature drops
below $V^{(2)}_\B$, thermal transitions among
different equivalence classes would cease to exist.
In each region of the space one of the equivalence classes would
be selected.   It is possible that
the universe forms a domain structure  such that  each domain
is in its own equivalence class.  Then, at the edge of each domain,
there forms a domain wall that connects two distinct equivalence
classes.  Such domain walls would be something exotic which could not be
described in the language of gauge theory alone.

The universe continues to expand and the temperature drops further.
When $V^{(2)}_\B > T > V^{(1)}_\B$,  the entire universe or each domain
in the universe remains in one equivalence class.  Wilson line phases
are thermally excited.  When $T$ drops below $V^{(1)}_\B$, the
Wilson line phases  settle into one of the minima of the effective
potential, which determines the low energy symmetry.
The Wilson line phases may happen to be trapped in a `false' vacuum
(a  local minimum) instead of the `true' vacuum (the global minimum).

The inflation may take place somewhere between the scales
$M_\GUT$ and $M_\SUSY$.  In this connection we recognize that
$V^{(1)}_\B \sim \sqrt{ M_\GUT M_\SUSY }$ defines the intermediate
scale.  The Wilson line phases themselves may serve as inflatons
at the scale $\sqrt{ M_\GUT M_\SUSY }$.
As shown in Ref.\ 24), the potential for the Wilson line phases
(or the extra-dimensional components of the gauge fields) takes
a special form which may be suited for natural
inflation.\cite{natural-inf} Such a scenario has been employed in Ref.\
33) to realize the quintessence scenario.  In our case, the
effective cosmological constant obtained is of $O(\sqrt{ M_\GUT
M_\SUSY})$,  not of the order of the cosmological constant observed
recently.

\section{Conclusions}

We have tackled the arbitrariness problem of the boundary conditions (BCs) in
$SU(N)$ gauge theory on the orbifold $S^1/Z_2$; i.e., we have
attempted to explain how one particular set of the boundary conditions 
is dynamically selected over  many other possibilities.
Two theories are equivalent  if they
are related by a BC-changing gauge transformation.
According to the Hosotani mechanism, the physical symmetry of each
equivalence class is uniquely determined by the dynamics of the Wilson line
phases. Hence the number of inequivalent  theories is equal to the number of
the equivalence classes of the BCs.
We have classified the equivalence classes of the BCs.  It is found 
 that
 each equivalence class always has a diagonal representative
$(P_0, P_1)$ and that the number of equivalence classes is $(N+1)^2$
in $SU(N)$ gauge theory.

Next, we have derived generic formulas for the one-loop effective potential
at the vanishing Wilson line phases.  These formulas can be applied to any
equivalence class.
It has been presumed that $(P_0^\sym, P_1^\sym)$ is diagonal, as
has been confirmed in many examples investigated to this date.
When applied to non-supersymmetric theory, 
there arises an intrinsic ambiguity in comparing the  energy densities in
two theories in different equivalence classes;
the difference between the vacuum energy densities is, in general, 
infinite, and therefore one cannot compare them.

The unambiguous comparison of the vacuum energy densities in two
theories in different equivalence classes becomes possible in
supersymmetric theories.
We have found that in the supersymmetric $SU(5)$
models with the Scherk-Schwarz supersymmetry breaking, the theory with
the BCs yielding the standard model symmetry can be in
the equivalence class with the lowest energy density,
though the low energy theory may not reproduce the minimal
supersymmetric standard model.
Possibilities to derive a low energy theory
with the standard model gauge group have been studied
particularly for the equivalence classes $[p; q,r;s] = [2; 3,0; 0]$ and 
$[p; q,r;s] = [2; 0,0; 3]$.

Further, we have discussed how particular BCs are selected 
in the cosmological evolution of the universe.
It is believed that the quantum tunneling transition rate at zero
temperature from one
equivalence class to another is probably negligibly small, though
the dynamics connecting such different equivalence classes 
are not  understood at all.
In one scenario,
the thermal transitions among different equivalence classes
cease to exist below a temperature whose magnitude is roughly
that of the   energy barrier separating  them.
The universe may or may not form a domain structure and settle into one of
the minima of the effective potential as the universe cools  further.

We would like to stress that the arbitrariness problem
has not been completely solved yet as
there  remains a degeneracy among the theories of the lowest energy density.
A new mechanism must be found to lift this degeneracy.
It is certainly  necessary to understand the dynamics in a more fundamental
theory in order to  determine the boundary conditions.

\vskip 1cm

\section*{Acknowledgments}
We would like to thank S. Odake  for helpful discussions.
N.\ H.\ and Y.\ K.\ thank the Yukawa Institute for Theoretical Physics
at Kyoto University, where a part of this work was done
during the YITP-Workshop-03-04 on ``Progress in Particle Physics''.
Y.\ H.\ would like to thank the Theoretical Physics  Department of the Fermi
National Accelerator
Laboratory for its hospitality where a part of the work was done.
Stimulating discussions there with Y.\ Nomura, W.\ Bardeen and C.\ Hill are
duly appreciated.  This work
was supported in part by  Scientific Grants from the Ministry of Education
and Science, Grant No.\ 13135215, Grant No.\ 13640284 (Y.H.), Grant No.\
14039207, Grant No.\ 14046208, Grant No.\ 14740164 (N.H.),
Grant No.\ 13135217 (Y.K.), Grant No.\ 15340078 (N.H., Y.H., Y.K.).

\appendix
\section{Diagonal representative of $(P_0, P_1)$}

The boundary condition matrices $P_0$ and $P_1$ are $N \times N$ hermitian,
unitary matrices.  As discussed in \S 2, two distinct sets, $(P_0,
P_1)$
and $(P_0', P_1')$, can be related by a large gauge transformation.
The two theories are said to be in the same equivalence class when
(\ref{BC4}) and (\ref{equiv1}) hold.  In this appendix we  prove
that in every equivalence class there is  at least one diagonal
$(P_0, P_1)$.

Through a global $SU(N)$ transformation, $P_0$ can be diagonalized as
\beeq
P_0 = 
\begin{pmatrix}
I_m \cr & -I_n
\end{pmatrix}
\next
P_1 = 
\begin{pmatrix}
A & C^\dagger\cr  C & B
\end{pmatrix} ~~,
\label{diagonal1}
\eneq
where $I_m$ is an $m\times m$ unit matrix and $m+n=N$.
$P_0$ still has $SU(m) \times SU(n)$ invariance. Utilizing this
invariance, one can diagonalize the hermitian matrices $A$ and $B$. Let us
write
\beeq
P_1 = 
\begin{pmatrix}
   a_1 &&&  & \vec c_1{}^\dagger \cr
   & \ddots &&  & \vdots \cr
   && a_m &  & \vec c_m{}^\dagger \cr
   &&&   b_1 \cr
   \vec c_1  &\cdots &\vec c_m &&\ddots \cr
   &&&  && b_n 
\end{pmatrix}
= 
\begin{pmatrix}
   a_1 &&&  &  \cr
   & \ddots &&  \vec d_1 &\cdots &\vec d_n \cr
   && a_m &   \cr
   &\vec d_1{}^\dagger &&   b_1 \cr
   &\vdots && &\ddots \cr
   &\vec d_n{}^\dagger && && b_n 
\end{pmatrix}  ~.
\label{diagonal2}
\eneq
As $P_1$ is unitary, we have
\beqn
&&\hskip -1cm
a_j^2 + \vec c_j{}^\dagger \vec c_j = 1 \next
b_j^2 + \vec d_j{}^\dagger \vec d_j = 1 ~~,  \cr
&&\hskip -1cm
\vec c_j{}^\dagger \vec c_k = \vec d_j{}^\dagger \vec d_k = 0
\quad \hbox{for } j \not= k ~.
\label{unitary1}
\eeqn

Let the rank of $C$ be $r$.  Only $r$ of the vectors $\vec c_j$ are
linearly independent. (\ref{unitary1}) implies that the other ($m-r$) of
the vectors
$\vec c_j$ identically vanish.  Similarly, only $r$ of the vectors $\vec
d_j$ are nonvanishing.  Through a rearrangement of the rows and columns, one
can bring
$P_1$ into the form
\beqn
&&\hskip -1cm
P_1 = 
\begin{pmatrix}
 \widetilde P_1 \cr &\hat I_{N-2r}  
\end{pmatrix}
 ~~,  \cr
\noalign{\kern 10pt}
&&\hskip -1cm
\widetilde P_1 = 
\begin{pmatrix}
a_1 &&&   \cr
& \ddots &&&\widetilde C^\dagger \cr
&& a_r & \cr
&&&  b_1 \cr
& \widetilde C & &&\ddots \cr
&&& &&b_r 
\end{pmatrix}  ~~.
\label{diagonal3}
\eeqn
Here, $\widetilde C$ is a square $r\times r$ matrix,  and $\hat I_{N-2r}$ is
a diagonal matrix whose diagonal elements are either $+1$ or $-1$.
Notice that
$(\widetilde P_1)^2 =I_{2r}$ implies
\beeq
(a_j + b_k) \, \widetilde C_{jk} = 0 ~~. \quad (1 \le j, k \le r) 
\label{unitary2}
\eneq
Making use of (\ref{unitary1}) and (\ref{unitary2}), one can reshuffle
rows and columns such that
\beqn
&&\hskip -1cm
\widetilde P_1 = 
\begin{pmatrix}
    \widetilde P_1^{(1)} \cr
    & \ddots \cr
    &&\widetilde P_1^{(t)} 
\end{pmatrix} ~~, \cr
\noalign{\kern 10pt}
&&\hskip -1cm
\widetilde P_1^{(l)} =
 \begin{pmatrix}
 a_{l}  I_{s_l} &C_{l}^\dagger \cr
             C_{l} &- a_{l}  I_{s_l} 
\end{pmatrix}   ~~, \cr
\noalign{\kern 10pt}
&&\hskip -.5cm
(s_1 + \cdots + s_t = r) ~~.
\label{diagonal4}
\eeqn

Now, consider the  submatrix $\widetilde P_1^{(l)}$.  It  is
also hermitian and unitary, which in particular implies that
$U^{(l)} = (1-a_l^2)^{-1/2}
C_l^\dagger$ is unitary; $U^{(l)\dagger} U^{(l)} = I_{s_l}$.
Through another global unitary transformation,  $\widetilde P_1^{(l)}$ is
brought into the canonical form:
\beeq
\begin{pmatrix}
 I_{s_l} \cr & U^{(l)} 
\end{pmatrix}
 ~\widetilde P_1^{(l)}~
\begin{pmatrix}
 I_{s_l} \cr & U^{(l)\dagger} 
\end{pmatrix}
=
\begin{pmatrix}
 a_l I_{s_l} & \sqrt{1-a_l^2} ~ I_{s_l} \cr
          \sqrt{1-a_l^2} ~ I_{s_l} & - a_l I_{s_l} 
\end{pmatrix} ~~.
\label{diagonal5}
\eneq
We note that $P_0$ remains invariant under this transformation.
The new $\widetilde P_1^{(l)}$ decomposes into $SU(2)$ submatrices.
In each subspace, we have
\beeq
\hat P_0=\tau_3 \next
\hat P_1 = \cos\theta \tau_3 + \sin\theta \tau_1 = e^{-i\theta \tau_2}
\tau_3
\next  \cos\theta = a_l ~.
\label{diagonal6}
\eneq
Finally, the $y$-dependent transformation
$\Omega(y) = e^{i(\theta y/2\pi R) \tau_2} $ in the subspace, with
(\ref{BC3}), transforms $(\hat P_0, \hat P_1)$ into $(\tau_3, \tau_3)$ in
the subspace. Hence,  the original general  $(P_0, P_1)$ has been
transformed into a diagonal one under a series of transformations.
This completes the proof.

\def\jnl#1#2#3#4{{#1}{\bf #2} (#4), #3}

\def\Zphys{{\em Z.\ Phys.} }
\def\jssc{{\em J.\ Solid State Chem.\ }}
\def\jpsJ{{\em J.\ Phys.\ Soc.\ Japan }}
\def\ptps{{\em Prog.\ Theoret.\ Phys.\ Suppl.\ }}
\def\PTP{{\em Prog.\ Theoret.\ Phys.\  }}

\def\JMP{{\em J. Math.\ Phys.} }
\def\NPB{{\em Nucl.\ Phys.} B}
\def\NP{{\em Nucl.\ Phys.} }
\def\PLB{{\em Phys.\ Lett.} B}
\def\PL{{\em Phys.\ Lett.} }
\def\PRL{\em Phys.\ Rev.\ Lett. }
\def\PRB{{\em Phys.\ Rev.} B}
\def\PRD{{\em Phys.\ Rev.} D}
\def\PRe{{\em Phys.\ Rep.} }
\def\AP{{\em Ann.\ Phys.\ (N.Y.)} }
\def\RMP{{\em Rev.\ Mod.\ Phys.} }
\def\ZPC{{\em Z.\ Phys.} C}
\def\SCI{\em Science}
\def\CMP{\em Comm.\ Math.\ Phys. }
\def\MPLA{{\em Mod.\ Phys.\ Lett.} A}
\def\IJMPA{{\em Int.\ J.\ Mod.\ Phys.} A}
\def\IJMPB{{\em Int.\ J.\ Mod.\ Phys.} B}
\def\EPJC{{\em Eur.\ Phys.\ J.} C}
\def\PR{{\em Phys.\ Rev.} }
\def\JHEP{{\em J.\ High Energy Phys.} }
\def\cmp{{\em Com.\ Math.\ Phys.}}
\def\JPA{{\em J.\  Phys.} A}
\def\CQG{\em Class.\ Quant.\ Grav. }
\def\ATMP{{\em Adv.\ Theoret.\ Math.\ Phys.} }
\def\ibid{{\em ibid.} }

\end{document}